\documentclass[namedreferences]{SolarPhysics}

\usepackage[optionalrh]{spr-sola-addons} 
\usepackage{graphicx}                    
\usepackage{url}                         

\newcommand{\aap}{{\it Astron. Astrophys.}}
\newcommand{\aaps}{{\it Astron. Astrophys. Suppl.}}

\newcommand{\apj}{{\it Astrophys. J.}}
\newcommand{\araa}{{\it Ann. Rev. Astron. Astrophys.}}

\newcommand{\nat}{{\it Nature}}

\newcommand{\solphys}{{\it Solar Phys.}}

\newcommand{\planss}{\it {Planet. Space Sci.\ }}
\newcommand{\lA}{{\bfseries\sffamily (A)}}
\newcommand{\lB}{{\bfseries\sffamily (B)}}
\newcommand{\lC}{{\bfseries\sffamily (C)}}

\begin{document}

\begin{article}

\begin{opening}

\title{Sub-terahertz, microwaves and 
high energy emissions during the 6 December 2006 flare, at 18:40 UT}

%
\author{Pierre~\surname{Kaufmann}$^{1,2}$\sep
       G\'erard~\surname{Trottet}$^{3}$\sep
       C.~Guillermo~\surname{Gim\'enez de Castro}$^{1}$\sep
       Jean-Pierre~\surname{Raulin}$^{1}$\sep
       S\"am~\surname{Krucker}$^{4}$\sep
       Albert~Y.~\surname{Shih}$^{4}$\sep
       Hugo~\surname{Levato}$^{5}$
       }

  \institute{$^{1}$ Centro de R\'adio-Astronomia e Astrof\'{\i}sica Mackenzie, 
                    Escola de Engenharia, Universidade Presbiteriana Mackenzie, 
                    Rua Consola\c{c}\~ao 896, 01302-907 S\~ao Paulo, SP, Brazil.
                    email:\url{kaufmann@craam.mackenzie.br}\\
             $^{2}$ Centro de Componentes Semicondutores,Universidade Estadual 
de Campinas, 
                    Campinas, SP, Brazil.\\
             $^{3}$ LESIA, Observatoire de Paris-Meudon, Meudon, France.\\
             $^{4}$ Space Sciences Laboratory, University of California 
Berkeley, Berkeley, CA, USA.\\
             $^{5}$ Complejo Astron\'omico El Leoncito, San Juan, Argentina.
             }

\begin{abstract}
The presence of a solar burst spectral component with flux density
increasing with frequency in the sub-terahertz range, spectrally separated
from the well-known microwave spectral component, bring new possibilities
to explore the flaring physical processes, both observational and
theoretical. The solar event of 6 December 2006, starting at about 18:30
UT, exhibited a particularly well-defined double spectral structure, with
the sub-THz spectral component detected at 212 and 405 GHz by SST and
microwaves (1-18 GHz) observed by the Owens Valley Solar Array
(OVSA). Emissions obtained by instruments in satellites are discussed with
emphasis to ultra-violet (UV) obtained by the Transition Region And Coronal
Explorer (TRACE), soft X-rays from the Geostationary Operational
Environmental Satellites (GOES) and X- and $\gamma$-rays from the Ramaty
High Energy Solar Spectroscopic Imager (RHESSI). The sub-THz impulsive
component had its closer temporal counterpart only in the higher energy X-
and $\gamma$-rays ranges. The spatial positions of the centers of emission
at 212 GHz for the first flux enhancement were clearly displaced by more
than one arc-minute from positions at the following phases. The observed
sub-THz fluxes and burst source plasma parameters were found difficult to
be reconciled to a purely thermal emission component. We discuss possible
mechanisms to explain the double spectral components at microwaves and in
the THz ranges.
\end{abstract}


\end{opening}

\section{Introduction}\label{sect:intro} 

Although radiative signatures of non thermal particles produced in flares
have been extensively studied, the physics of particle acceleration at the
Sun is not yet fully established. The presence of an independent flare
terahertz spectral emission component, simultaneously to the well known
component at microwaves, brought new challenges and possibly new clues to
understand the nature of the flare primary accelerator
\cite{Kaufmannetal:2004}. The vast majority of microwave bursts exhibits
typical spectra with maximum fluxes in the range 5-20 GHz. Fewer
observations carried out at higher frequencies, up to 100 GHz, have shown
uncommon events exhibiting fluxes increasing with frequency, some showing
complex spectral shapes, and inflections somewhere between 30-70 GHz, and
other flattening at higher frequencies
\cite{Croom:1970,Croom:1971b,Shimabukuro:1970,Cogdell:1972,Akabaneetal:1973,Kaufmannetal:1985,Ramatyetal:1994,Kleinetal:1999,Bastianetal:2007,Altyntsevetal:2008}.
Recent observations of intensity increases with frequency at sub-THz
frequencies provide evidence for a new spectral component clearly separated
from the well known emission spectrum at microwaves, displaying a
double-structure in the microwave - submillimeter range of wavelengths. \\

We will describe and discuss the solar flare of 6 December 2006, starting
at about 18:30 UT that exhibited a distinct THz spectral component
throughout the event duration as observed by the solar submillimeter
telescope (SST) at 212 and 405 GHz \cite{Kaufmannetal:2007} until 18:55
UT. Minutes later, the OVSA (Owens Valley Solar Array) has detected the
largest decimeter narrow band spikes ever observed \cite{Gary:2008}. SST
observations were compared to decimeter to microwaves obtained by the Owens
Valley Solar Array and to X- to $\gamma$-rays detected by RHESSI satellite.
These observations were complemented by TRACE satellite UV movies. The
event corresponds to a 3B flare on AR 0930, S06E63, starting at 18:32 UT,
maximum at 18:45 UT, and at 21:35 UT, with soft X-ray GOES class X6.5.

\section{Multiple wavelength time profiles and spectra}\label{sect:spectra}

\begin{figure} 
\centerline{\includegraphics[width=10cm]{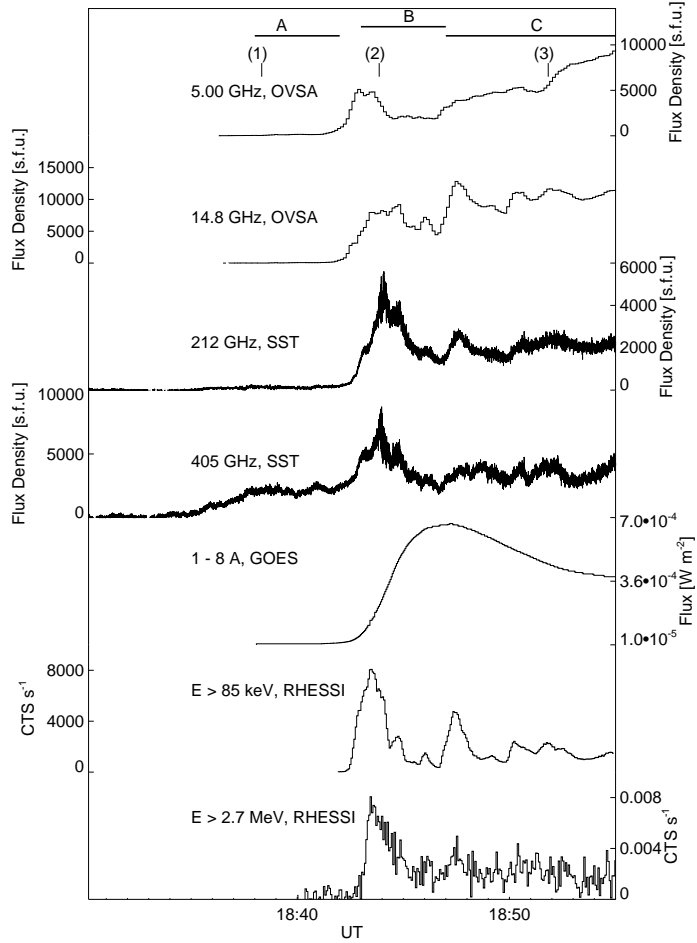}}
\caption{Time profiles for the 6 December 2006 solar burst. From top to
bottom: two microwaves from OVSA; the two SST frequencies; GOES soft
X-rays; RHESSI at hard X-rays and at $\gamma$-rays. Labels \lA, \lB\ and
\lC\ are the time intervals for which the positions were taken for the 212
GHz burst centroids of emission, shown in Figure
\ref{fig:sst-anttemp}. Labels 1, 2 and 3 are the times when the burst
sub-THz and microwave spectra were taken, shown in Figure
\ref{fig:mw-spectra}.}\label{fig:timeprofiles}
\end{figure}

We show in Figure \ref{fig:timeprofiles} the burst intensity time profiles
at two microwave frequencies, submillimeter-waves, soft, hard X-rays and
$\gamma$-rays. Despite of good clear-sky conditions at SST El Leoncito site
the atmosphere measured attenuation was high.  It was measured at 15:30 UT,
before the burst, providing optical depths $\tau$ of 0.35 and 2.6 nepers at
212 and 405 GHz, respectively. At a mean elevation angle of 55$^\circ$ the
correction factors to correct antenna noise temperatures were of 1.5 and 24
respectively. There were indications of transmission changes after 18:55 UT
when the 405 GHz data corrections became inconsistent.  Since0811.3560 there were no
other atmosphere transmission measurements, the analysis was interrupted.
Small changes in large values of $\tau$, as was the case for 405 GHz, may
bring significant changes in correction factors, which cannot be taken into
account here. The labels at the top of Figure \ref{fig:timeprofiles} refer
to: the precursor-like first enhancement \lA, the maximum impulsive-like
phase \lB\ and following time structures \lC. In phase \lB\ the sub-THz
time profile compares to the bulk of impulsive emissions at microwaves, $>$
2.7 MeV $\gamma$-rays and to a lesser extent to $>$ 85 keV hard X-rays.\\

\begin{figure} 
\centerline{\includegraphics[width=\textwidth]{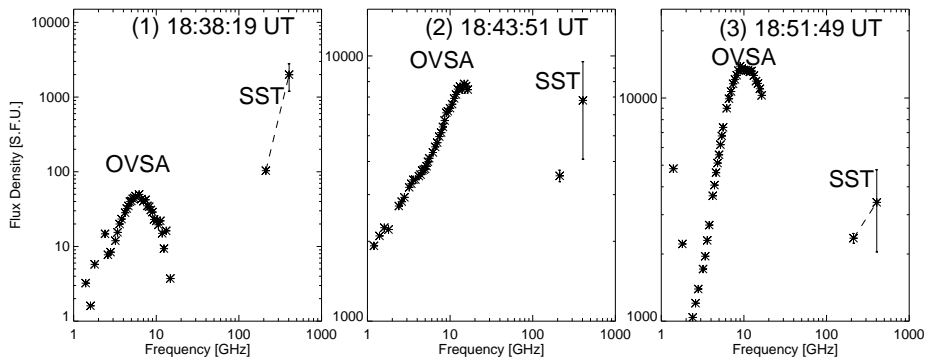}}
\caption{Complete burst spectra, from decimeter to submillimeter waves, for
the burst times 1-3 labeled at the top of Figure \ref{fig:timeprofiles}.
Bars refer to an arbitrary uncertainty assumption of 10\% changes in the
optical depths.  The presence of the sub-THz component is particularly well
defined for the precursor-like structure, the impulsive phase, and
suggested for the following phase.}\label{fig:mw-spectra}
\end{figure}

\begin{figure} 
\centerline{\includegraphics[width=12cm]{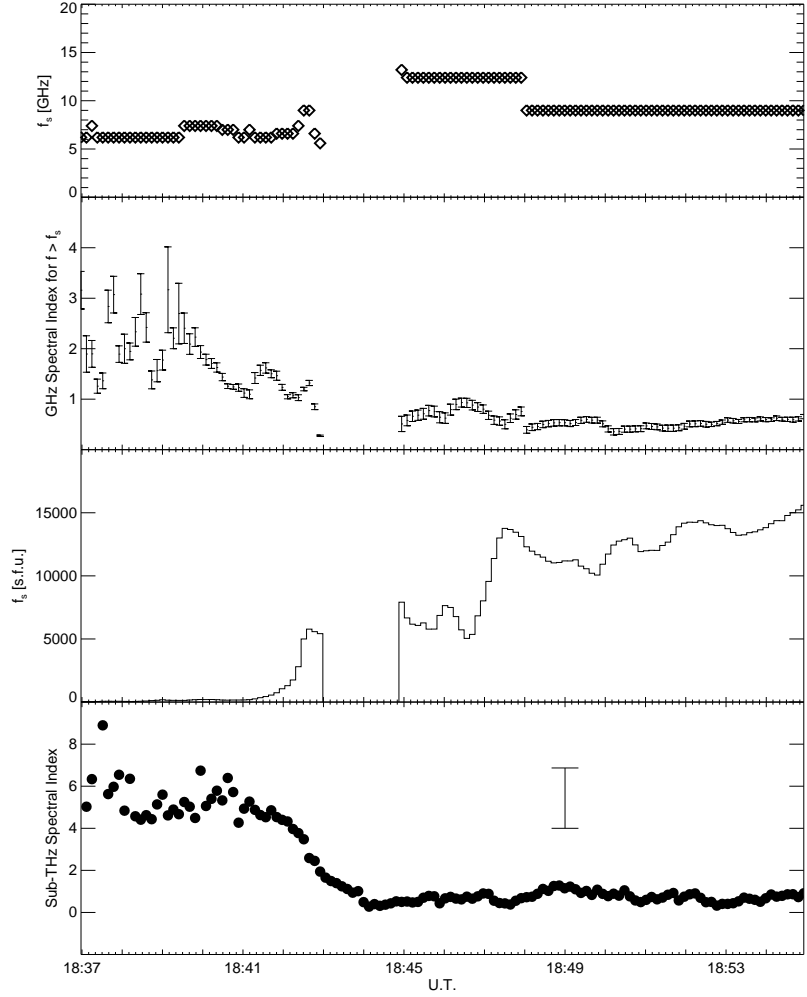}}
\caption{From top to bottom, time variation of: the spectral turnover 
frequency $f_s$ at microwaves derived from OVSA observations; 
the microwave spectral index $\delta$, defined as flux $\propto f^{-\delta}$
for $18 \ \mathrm{GHz} \ge f \ge f_s$ ; of the microwave flux at the 
turnover frequency $f_s$. Missing data correspond to time intervals where $f_s$ 
can be higher than the maximum OVSA frequency (18 GHz). Bottom:
time variation of the sub-THz spectral index (see text). The bar represents
the fixed uncertainty on the 405 GHz flux correction due to atmosphere 
transmission. }
\label{fig:turnover}
\end{figure}

Examples for the 212 and 405 GHz spectra are shown in Figure
\ref{fig:mw-spectra} sampled over 8.1 seconds averages (to be comparable to
the OVSA time resolution) on the times labeled 1-3 at the top of Figure
1. The burst spectra exhibited the two distinct components throughout the
whole event duration, one peaking at microwaves, as derived from Owens
Valley Solar Array, and another sub-THz component increasing with
frequency, as observed by the SST.  The uncertainty bars indicate the
extreme limits set empirically for a conservative assumption of 10\%
changes in the optical depths. The sub-THz spectral component together with
the independent microwave component is strikingly evident at the
precursor-like phase \lA, being present at the impulsive phase \lB, and
suggested for the following phase \lC.  The OVSA microwave spectra for the
three phases are shown in Figure \ref{fig:mw-spectra}. They are usually
attributed to gyro-synchrotron emission by mildly relativistic electrons
\cite{Dulk:1985}. The maximum emission turnover frequencies change with
time, as it has been known for other bursts
\cite{Croom:1971a,Roy:1979,Nitaetal:2004}.  Figure \ref{fig:turnover} shows
the time variation of microwave turnover frequencies $f_s$ (which in some
cases cannot be unambiguously determined because apparently it exceeded the
maximum OVSA limit), the GHz spectral index $\delta$ (defined as flux
$\propto f^{-\delta}$) for $18 \ \mathrm{GHz} \ge f \ge f_s$ and the GHz
flux at $f_s$.  In the bottom of the same figure we show the sub-THz
spectral index, calculated as $\delta_{sm} =
\log[S(405)/S(212)]/\log[405/212]$, with $S(405)$ and $S(212)$ the flux
density at 405 and 212 GHz respectively. Rapid superimposed time structures
were present at both sub-THz frequencies and will be discussed in a
separate paper.\\

\section{Positions, sizes and fluxes}\label{sect:flux}

The positions of burst centers of emission can be determined at 212 GHz by
comparing the relative intensities (corrected antenna temperatures) of the
SST three partially overlapping beams
\cite{Georgesetal:1989,Costaetal:1995,Castroetal:1999}.  Beam and source
shapes are approximated to Gaussians. The source's spatial structures,
however, cannot be resolved within the diffraction limit set by the beam
angular sizes (4 arc-minutes).  A crude estimation of the angular extent
occupied by the source with respect to the beam can be obtained by means of
the contrast parameter: a ratio between the three corrected antenna
temperatures. When the source size is small compared to the beam size, the
contrast should be high, and the contrary happens for an extended source.\\

\begin{figure} 
\centerline{\includegraphics[width=\textwidth]{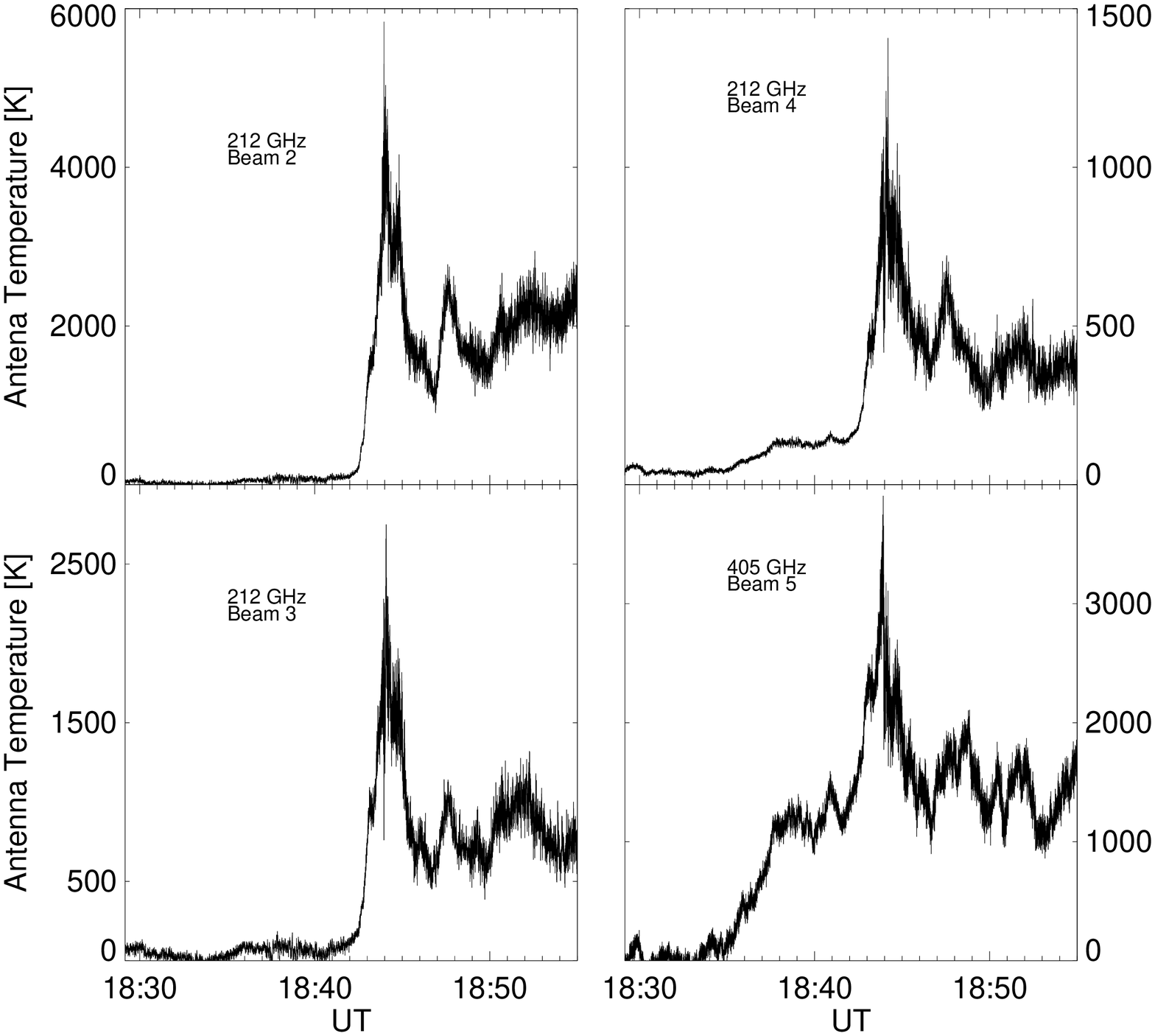}}
\caption{Antenna noise temperatures, corrected for atmosphere transmission,
for the SST receivers whose beams are on the burst sources, 5 (405 GHz) 2,
3 and 4 (212 GHz). The beam positions on the solar disk are shown in Figure
\ref{fig:sst-positions}.  Note that the temperature scales are different
for the different panels. The precursor-like and impulsive time structures
are well correlated. The following phase 212 GHz fluctuations are poorly
related to the 405 GHz time variations, which might receive considerable
influence from atmosphere transmission variations.}\label{fig:sst-anttemp}
\end{figure}

\begin{figure} 
\centerline{\includegraphics[width=0.7\textwidth]{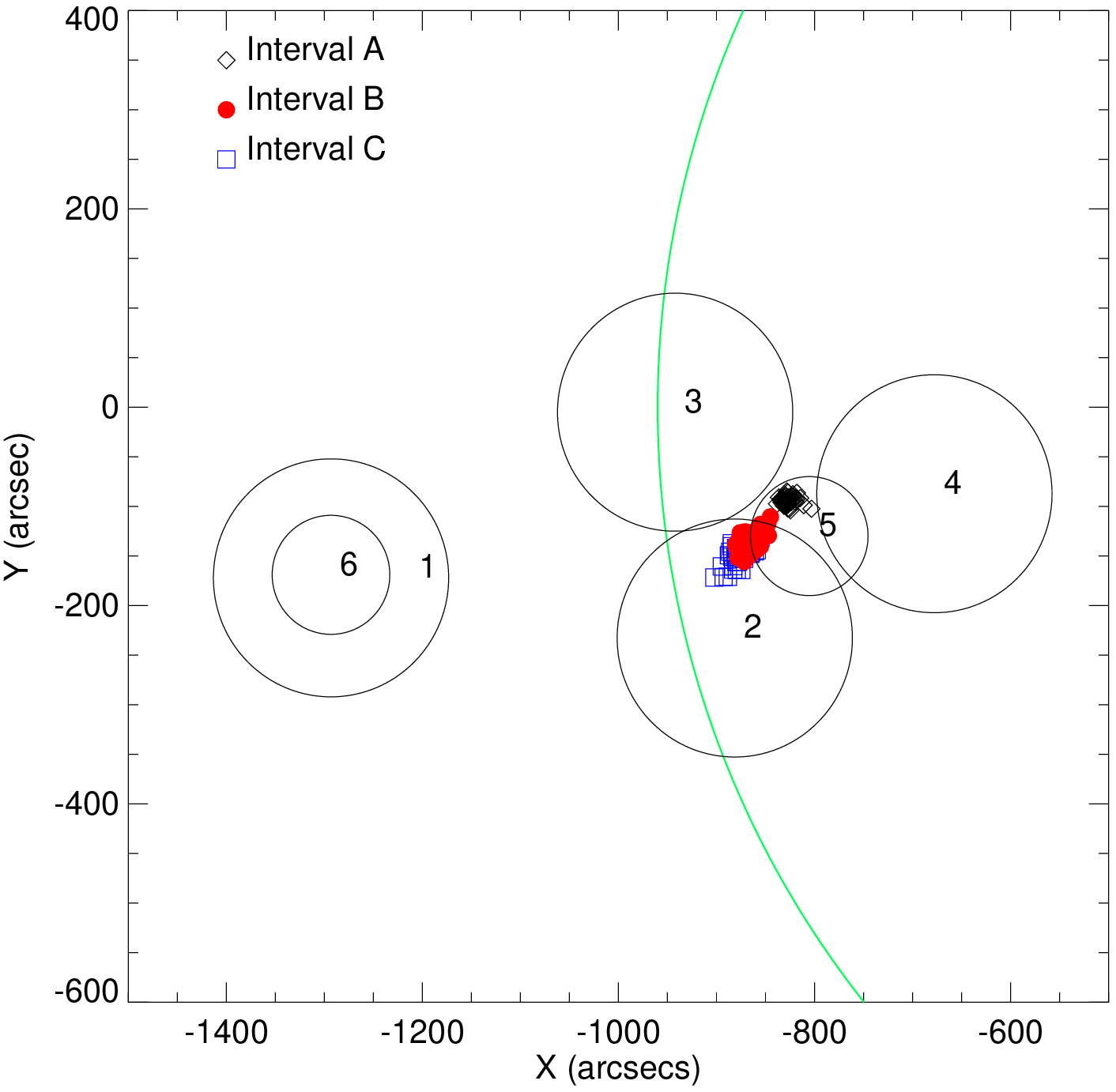}}
\caption{The positions for the 212 GHz burst centroids of emission,
averaged every second along the time intervals indicated in the top of
Figure 1, for phases \lA\ (precursor),\lB\ (impulsive) and \lC\
(following). The precursor position is clearly displaced by about one
arc-minute from the remaining burst phases, which are close by about 30
arc-seconds to each other. The six SST antenna beams are shown with respect
to the solar disk at the time of the burst. They are considerably larger
than the emitting source centroid displacements. For the precursor-like
phase the 212 GHz centroid of emission is closer to beams 3 and 4 producing
larger corrected antenna temperatures, shown in Figure
\ref{fig:sst-anttemp}.}\label{fig:sst-positions}
\end{figure}

Figure \ref{fig:sst-anttemp} shows the antenna noise temperature time
profiles, corrected for atmospheric transmission, for the beams spatially
close to the burst location.  There is a good temporal agreement during
phases \lA\ and \lB. 405 GHz time structures at phase \lC\ (beam 5) are not
always related to the 212 GHz time structures (beams 2, 3 and 4). This
might be attributed to the enhancement of atmosphere transmission
variations, substantially larger at 405 GHz. The 212 GHz emission centroid
positions shown in Figure \ref{fig:sst-positions} have been estimated from
40 ms data and averaged over one second for intervals labeled \lA, \lB\ and
\lC\ in Figure \ref{fig:timeprofiles} (top).  The sizes and positions for
the six SST beams are also shown, with respect to the solar limb at the
time of the burst. Since December 2006, the SST pointing model has provided
an absolute position accuracy of 10 arc-seconds r.m.s. for all beams
\cite{Wallace:2006}. The positions inferred for precursor-like burst
(interval \lA) are clearly distinct by about one arc-minute from the
positions of the 212 GHz emitting region derived during phases \lB\ and
phase \lC. \\

The observed contrasts between the beam outputs at 212 GHz indicate that
the source size was at most of the order of the antenna beam size
throughout the whole event. Since just one beam is available at 405 GHz no
position for emission centers or burst size could be estimated at this
frequency. To determine the flux densities, we first compare the corrected
antenna temperatures for the three partially overlapping beams to determine
the direction for the maximum emission. The maximum antenna temperature is
calculated assuming Gaussian shape for the three beams. The c0811.3560alculated
fluxes are for sources placed at the direction of maximum emission. They
are obtained from the well known relationship for point sources (still
valid for sources small or comparable to the half-power beam size),
\begin{equation} S = \frac{2 k T_a}{A_e}\ \ ,\end{equation}
where $k$ is the Boltzmann constant, $T_a$ the antenna temperature and
$A_e$ the antenna effective area (0.35 m$^2$ at 212 GHz and 0.18 m$^2$ at
405 GHz at the time these measurements were obtained). At 212 GHz the
antenna temperature averaged over 8 seconds interval around the burst
maximum (at 18 43:51 UT) was $T_a \sim 4900$ K. This corresponds to a flux
density of about 3800 s.f.u. at 212 GHz (1 s.f.u = $10^{-22}$ W m$^{-2}$
Hz$^{-1}$). The 405 GHz flux densities are similarly calculated by assuming
the source has the same position as the 212 GHz source. The mean 405 GHz
peak temperature was $T_a \sim 4500$ K corresponding to a flux of 7000
s.f.u. The two temperatures are comparable suggesting emission from an
optically-thick source at the two sub-THz frequencies.\\

The burst fluxes refer to Gaussian-shaped sources with equivalent sizes of
the order or smaller than the beam sizes. No estimates can be directly
given for actual brightness temperatures of smaller sources that might be
scattered and unresolved over the beam extension. \\

\begin{figure}
\centerline{\includegraphics[width=0.48\textwidth]{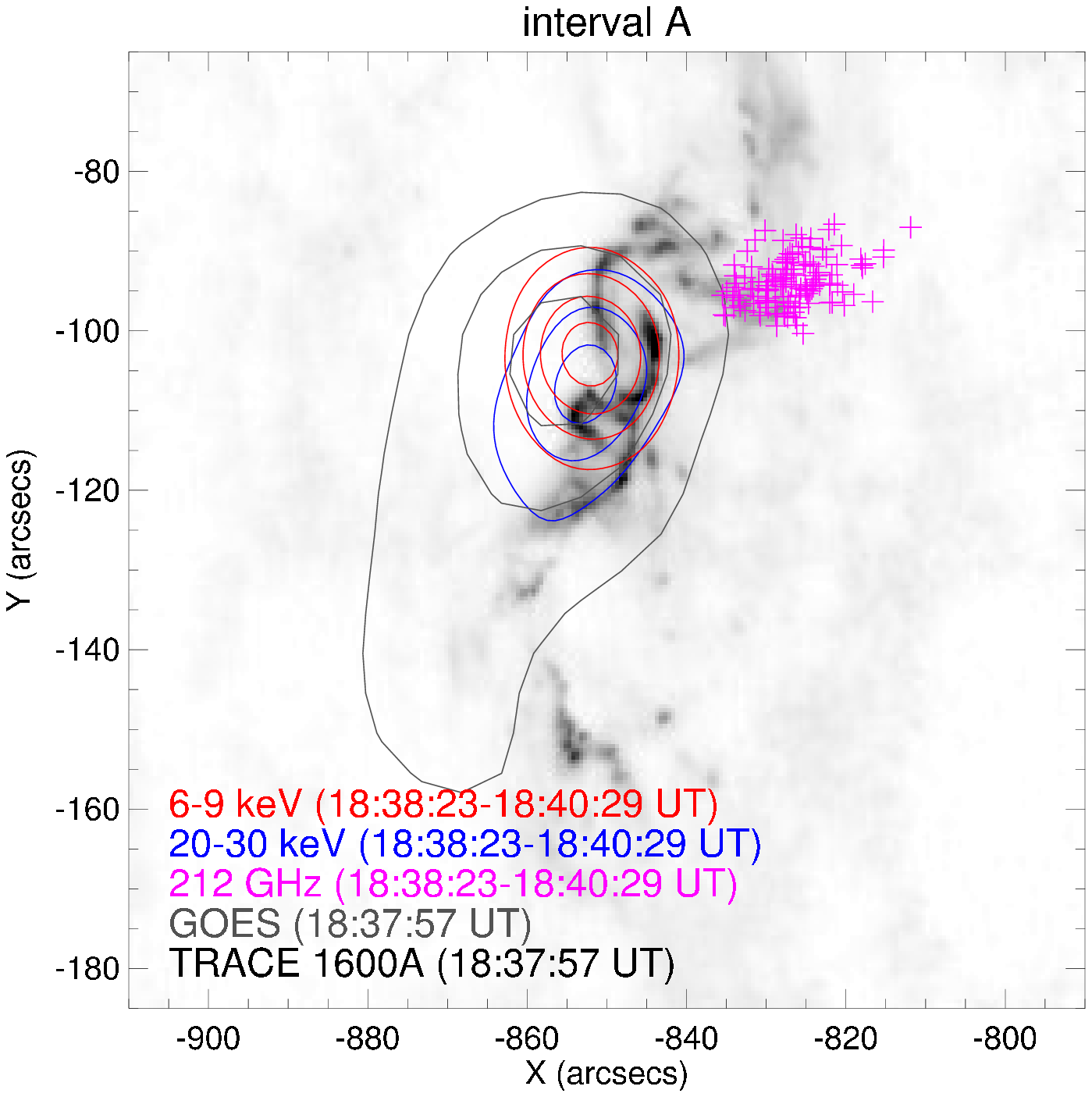} \quad
\includegraphics[width=0.48\textwidth]{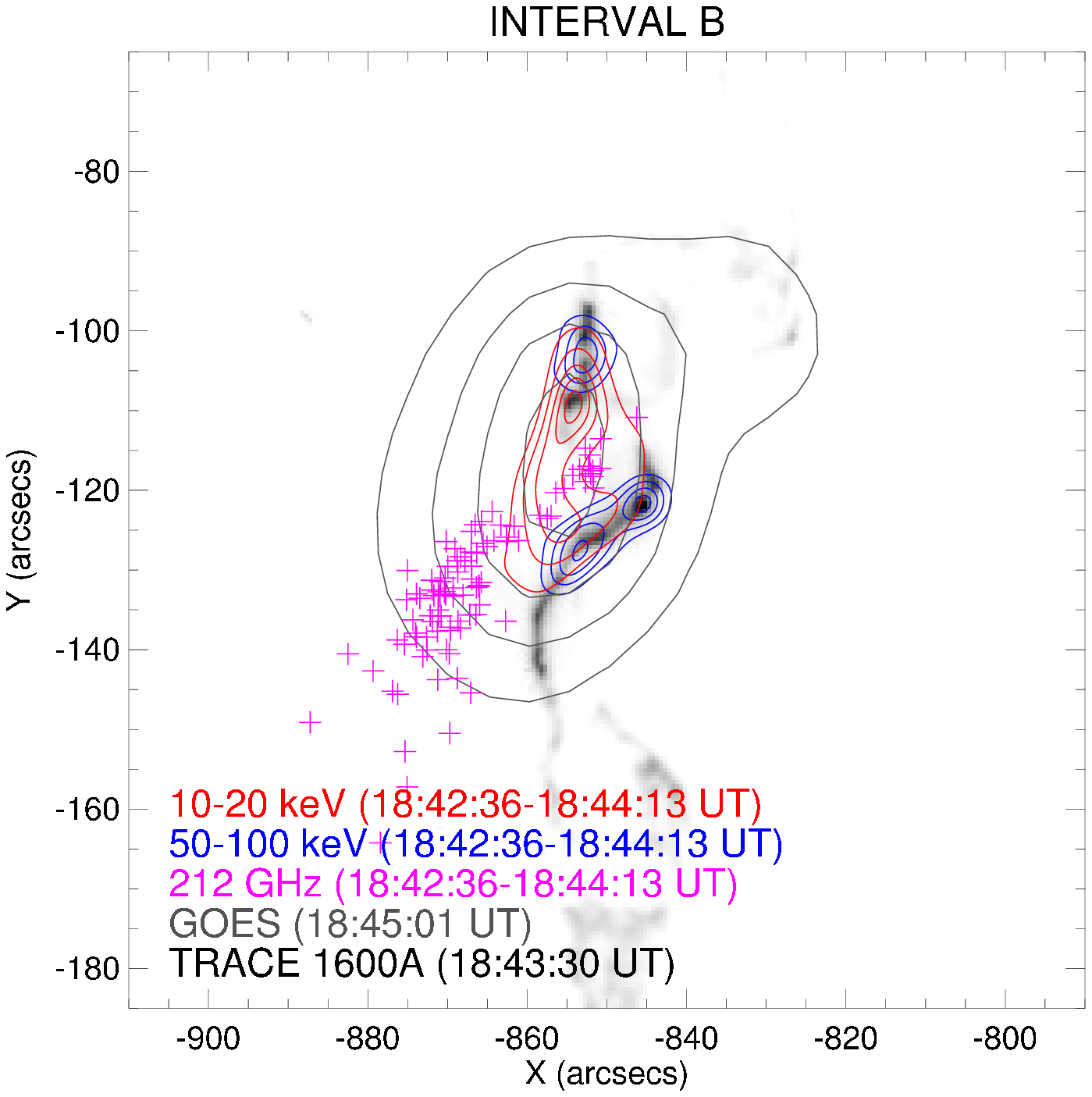}}
\centerline{\includegraphics[width=0.48\textwidth]{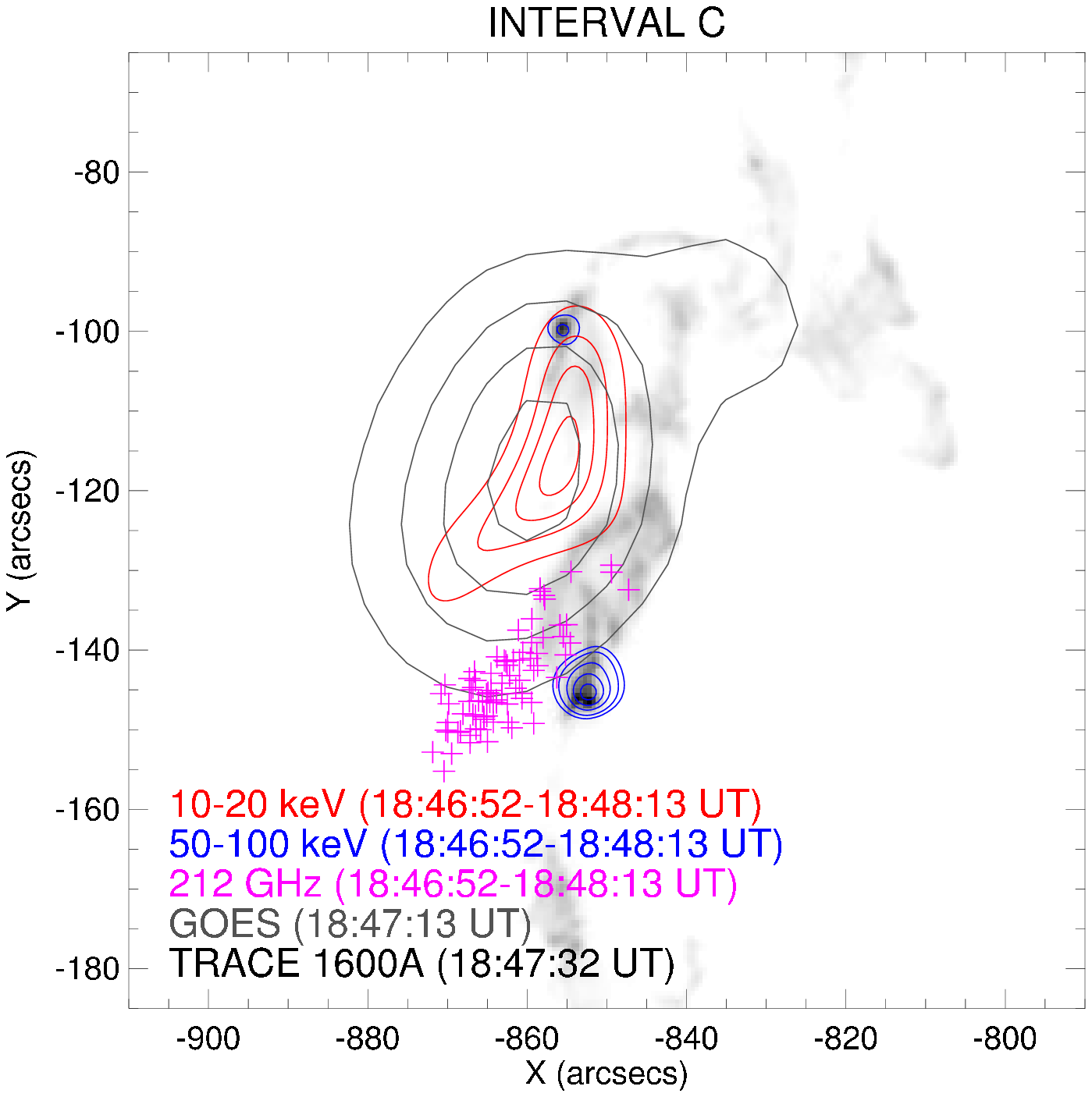}}
\caption{
RHESSI hard X-ray (thermal emission in red and non-thermal in blue
contours), GOES soft X-ray (thin black contours), and TRACE UV images 
background continuum gray-black areas during phase \lA, \lB, and
\lC (see Figure \ref{fig:timeprofiles}). 212 GHz centroid positions 
(same as in Figure \ref{fig:sst-positions}) are shown as magenta 
crosses for comparison.}\label{fig:hxr-positions}
\end{figure}

Figure \ref{fig:hxr-positions} shows emission contours at UV, soft- and
hard X-rays taken along time intervals within the three burst phases (see
labels at the top of the Figure). The precursor-like phase \lA\ was
detectable only at softer RHESSI X-ray energies ($>$ 6 and $>$ 20 keV). The
source is located about 20 arc-seconds east from the 212 GHz centroid of
emission at interval \lA. Two small RHESSI hard X-ray source ($>$ 50 keV)
appear at intervals \lB\ and \lC.  The southern hard X-ray source displaces
further south by about 20 arc-seconds from phase \lB\ to phase \lC. The 212
GHz emission centroids at phases \lB\ and \lC\ are displaced in the SE
direction by almost 1 arc-min.  There is a suggested centroid concentration
at phase \lC, displaced and close to the general hard X-Ray source
position.  The background UV TRACE images indicate complex spatial
structures.  Since the 212 GHz beam angular sizes are larger than the frame
angular sizes, they may subtend a number of unresolved sub THz sources.\\

\section{Discussion}\label{sect:discussion}

The 6 December 2006 flare exhibited the THz spectral component throughout
the event duration. It is particularly evident since the early phase \lA,
previous to the impulsive enhancement. The spectrally separated microwaves
component was always present, exhibiting a turnover frequency at about 6-7
GHz prior to the impulsive phase \lA, when it raised to about 15-13 GHz
\lB, staying at about 10 GHz during the remaining time.\\

\subsection{Thermal interpretation}\label{sect:thermal}

From GOES observations we derived the mean emission measure $EM=10^{50}$
cm$^{-3}$ and temperature $T=2.4 \ 10^7$ K (calculated after
\opencite{Garcia:1994} and \opencite{Feldmanetal:1995}) during the interval
1843:30 -- 1845 UT, which corresponds to the maximum emission at the
sub-THz frequencies.  Images from the SXI on GOES-12 revealed a soft-X ray
flaring size of approximately 1.5 arc min. With these values, the maximum
expected free-free emission (e.g. \opencite{Dulk:1985}) of a homogeneous
source is $\sim $ 100 s.f.u. with a turnover frequency around 10 GHz. It is
almost two orders of magnitude smaller than the observed radio fluxes.
Nonetheless, we can investigate the necessary conditions that a thermal
source should fulfill to account for the increase in flux above 200 GHz.\\

The optical depth $\tau_{ff}$ for free-free emission
(e.g. \opencite{OhkiHudson:1975}) is given by
\begin{equation} \tau_{ff} \simeq 0.19 \frac{N^2 H}{T^{3/2}\nu^2} \ \ ,\end{equation} 
where $N$ is the electron density, $T$ the plasma temperature, $H$ is
the source length along the line of sight and can be considered $H \simeq
D_s$ with $D_s$ the source scale size and $\nu$ the frequency. We may
try to reconcile this brightness to the observed antenna temperatures
from the well known approximation:
\begin{equation}T_a \simeq T_b \frac{\Omega_s}{\Omega_a} \ \ , \end{equation}
where $\Omega_s$ is the source solid angle size and $\Omega_a$ the antenna 
beam solid angle, can be approximately written as:
\begin{equation}T_a \simeq  T_b \left ( \frac{D_s}{D_a}\right )^2 \ \ , 
\end{equation}
where $D_a$ is the projected size on the Sun of  the antenna beam. 
Equating (4) with (2) we obtain
\begin{equation} N^2 \simeq 5.26 \ \tau_{ff} \ \frac{ T_a^{3/2} \ D_a^3 \ \nu^2}{D^4_s} 
\ \ .\end{equation}\\

The interpretations of the 212 and 405 GHz emission in terms of optically
thick free-free radiation ($\tau_{ff} >> 1$) implies $T_s=T_b$, $T_s$ being
the temperature of an isothermal radio source. For $D_s < 15$ arc-seconds,
from equation (4) we get $T_s > 10^6$ K that is temperatures to which the
GOES instrument is sensitive. However, EM is more than three orders of
magnitude larger than that obtained from GOES. Thus, like for other events
exhibiting the THz component, the thermal interpretation is not consistent
with a compact source
\cite{Kaufmannetal:2004,KaufmannRaulin:2006,Silvaetal:2007,Trottetetal:2008}.\\

\subsection{Non-thermal interpretation}\label{sect:nonthermal}

If the 212 and 405 GHz sources are small, non-thermal mechanisms by high
energy particles (electrons and ions) have to be considered to produce the
emissions observed during the 6 December 2006.  The optically thick
gyrosynchrotron emission and the free-free absorption might be taken in
consideration to account for the observed flux increase with frequency.
Assuming, for example, a burst cylindrical volume with $\sim 1$ arc-second
diameter and $\sim 1$ arc-second height, on a plasma medium with ambient
density of $\sim 10^{12}$ cm$^{-3}$ and a temperature of a few $10^5$ K,
implies in a moderate $\tau_{ff}(200 \ \mathrm{GHz}) \sim 2$. If the
sub-millimeter flaring area has a magnetic field $B \sim 2000$ G, and the
non-thermal electrons a mean density $\sim 10^{10}$ cm$^{-3}$ for energies
$E \ge E_\circ = 25$ keV the self-absorption will also contribute with the
attenuation of the microwaves while still producing enough emission at the
sub-THz domain. On the other hand, it can be shown that the effectiveness
of the Razin supression at 200 GHz requires a high medium density where
free-free absorption dominates. Of course the two sub-THz frequencies data
points for the 6 December 2006 burst are not sufficient to define the
complete synchrotron spectrum and the various parameters involved to shape
it. \\

A challenging explanation is needed for the "double spectral" emission,
with one microwave component peaking between 6 and 15 GHz and another
spectrally independent component peaking somewhere in the THz range. One
explanation might conceive independent accelerators at the flaring source,
producing different energy electrons nearly simultaneously, emitting at two
spectral components. However the observations available are too limited to
bring any favorable evidence for this possibility. Another scenario
suggested by \inlinecite{WildSmerd:1972} might be adopted, placing a single
accelerator closer to a single polarity foot-point injecting electrons into
a magnetic morphology split into two separate loops, one low altitude with
stronger field, another weaker field, higher above the solar surface,
originating the two synchrotron spectral components.  Although this
possibility was suggested in the discussion of other bursts of this class
\cite{Silvaetal:2007,Trottetetal:2008,Cristianietal:2008}, it requires a
number of selective free assumptions for the emissions not sufficiently
constrained by the existing observations.\\

Synchrotron emission from high energy positrons was first proposed by
\inlinecite{LingenfelterRamaty:1967} to account for microwaves. More
recently, \inlinecite{Trottet:2006} suggested that positrons from charged
pions would be an attractive possibility to explain the THz
component. Indeed, for one event, \inlinecite{Trottetetal:2008} showed
that: (i) the 210 GHz impulsive phase emission started simultaneously with
the pion production and; (ii) the 210 GHz emitting source coincided with a
region of ion interactions, distinct from regions of electron
interactions. Such an interpretation cannot be quantitatively studied in
the case of the 6 December 2006 flare because the $\gamma$-ray emission
from this flare was only measured below 17 MeV by RHESSI. This is a too low
energy to estimate the relative contributions of pions and electrons to the
$>$ 10 MeV $\gamma$-ray continuum (cf \opencite{Vilmeretal:2003} and
references therein). Nevertheless, it should be noted that, for similar
events, the number of positrons needed to explain the sub-THz emission was
found much larger than that derived from $\gamma$-ray measurements
\cite{Silvaetal:2007,Trottetetal:2008}. We should note, however, that there
is a theoretical possibility for enhanced positron production by
proton-proton interactions by the Drell-Yan process
\cite{Szpiegeletal:2007}. \\

Another suggestion was given by simulations of fully relativistic electron
beams propagating into high density and high magnetic field medium,
generating Langmuir waves producing strong backward e\-mis\-sions with
intensities larger at higher sub-THz frequencies as observed
\cite{Sakaietal:2006}. The same beam, with energies of about 2 MeV, might
also produce the 10-20 GHz microwaves as observed. This possibility was
also simulated using proton beams \cite{SakaiNagasugi:2007}. However, these
simulations have not taken into account the importance of absorption by the
surrounding dense medium. \\

Another possibility to explain a double-peak spectrum in the
mi\-cro\-wa\-ve-sub\-mil\-li\-me\-ter range is to consider particle-wave
instabilities in electron beams accelerated to ultra-relativistic energies,
known as "microbunching".  The THz spectral component is attributed to
incoherent synchrotron radiation (ISR) emitted by ultra-relativistic beams
of electrons. The beams are bunched when traversing inhomogeneous magnetic
field structures, which are known to be common to sunspots
\cite{Sturrock:1987,Antiochos:1998,Zhang:2005}. The GHz spectral component
arises from the broadband coherent synchrotron radiation (CSR) produced by
a wave-particle instability, as a result from the bunching of the high
energy accelerated electron beams. The mechanism that produce intense
broadband CSR was known for long time \cite{Nodvicketal:1954},but only
recently recognized in laboratory accelerators
\cite{Williams:2002,Carretal:2002,Byrdetal:2002}.
\inlinecite{KaufmannRaulin:2006} and \inlinecite{Klopf:2008} discussed the
relevance of this process for solar flares.  The mechanism is highly
efficient. The total power emitted by a bunch is proportional to the total
number of electrons in the beam emitting the observed ISR peaking in the
THz range, according to the standard interpretation, while the CSR peaking
in the GHz range is proportional to the square of the same number times a
form factor ($0\le f \le 1$, \opencite{Nodvicketal:1954}). This means that
only a small fraction of the accelerated electrons satisfying this
condition are needed to account for the observed microwaves. \\

The same electron beams producing the ISR may eventually collide into
denser regions in the solar atmosphere producing X- and $\gamma$-rays by
bremsstrahlung. It has been shown that the ISR spectral contribution to the
observed X- and $\gamma$-rays might also become significant, although not a
necessary condition \cite{KaufmannRaulin:2006}. The number of electrons
required to produce the hard X- and $\gamma$-rays emissions should be
compared to the number of electrons needed to produce the ISR spectrum in
the THz range (rather than at microwaves as is usually done).  This
approach may bring a possible explanation to the electron number
discrepancy, also known as the "electron number paradox"
\cite{BrownMelrose:1977,Kai:1986}.\\

\section{Concluding remarks}

The sub-THz emitting region could not be spatially resolved during the
6 December 2006 event.  Therefore, a thermal interpretation of the radio
emission above 200 GHz cannot be discarded if we assume a single source
whose size is larger than 15 arc-seconds. However, the rather good
similarity between hard X-ray and sub-THz time profiles and the fact that
hard X-rays are emitted in discrete sources with diameters of the order of
10 arc-seconds, suggest that the sub-THz radiation was produced in several
non-thermal sources. If this is the case, the emission, which should peak
in the THz region, might be attributed to incoherent synchrotron radiation
(ISR) from ultra relativistic electrons or positrons.\\

A non-thermal interpretation brings questions on the nature of the
microwave spectra observed together with the sub-THz
emission. Possibilities discussed above suggest that electrons are injected
into different magnetic arches of different magnetic field strengths and
electron densities. in order to produce the two separate synchrotron
spectral components.  The Razin effect might play a role to suppress the
optically thin synchrotron emission below 400 GHz. \\

The microwave emission might be at least partially produced by another
mechanism recently recognized in laboratory accelerators. Indeed, intense
broadband coherent synchrotron radiation (CSR) in the microwave range may
be produced as the result of the bunching of high energy electron
beams. These beams also produce the incoherent synchrotron radiation (ISR)
in the THz range.  The electron beams producing the ISR may eventually
collide into denser regions where they may contribute to the X- and
$\gamma$-ray bremsstrahlung continuum. It should also be noted that
additional contribution to hard X-rays and $\gamma$-rays might come from the
high frequency part of the ISR spectrum.\\
	 
Progresses in this field of research require measurements in the unexplored
THz range which are crucial for the understanding of the relative
importance of the mechanisms discussed in this study. New experiments are
currently being considered to observe solar flares from ground and
space. The most advanced is the DESIR (DEtection of Solar eruptive Infrared
Radiation) experiment for the France-China satellite SMESE (SMall Explorer
for the study of Solar Eruptions) \cite{Vialetal:2008}.

\begin{acks}
The authors acknowledge D.E. Gary for making available the OVSA microwave
data and the the support given to SST operations by the Complejo
Astronomico El Leoncito engineers A. Marun, P. Pereyra, R. Godoy and
G. Fernandez.  This research was partially supported by Brazilian agencies
FAPESP, CNPq, MackPesquisa, Argentina agency CONICET and France agency
CNRS.
\end{acks}

\end{article} 
\end{document}